\documentclass[journal]{IEEEtran}
\usepackage{algorithmic}
\usepackage{amsmath}
\usepackage{amsfonts}
\usepackage[pdftex]{graphicx}
\graphicspath{{fig/}}
\usepackage{threeparttable}
\usepackage{booktabs}  
\usepackage[linesnumbered, ruled, vlined]{algorithm2e}
\usepackage{cite}
\usepackage{hyperref}

\begin{document}
	
\title{Open-Access Data and Toolbox for Tracking COVID-19 Impact on Power Systems
\thanks{G.~Ruan, Z.~Yu, S.~Pu, S.~Zhou, H.~Zhong, Q.~Xia, and C.~Kang are with the State Key Lab of Power Systems, Department of Electrical Engineering, Tsinghua University, Beijing 100084, China.}
\thanks{L.~Xie is with the Department of Electrical and Computer Engineering, Texas A\&M University, College Station, TX 77843, USA.}
\thanks{The work of L. Xie is supported in part by NSF ECCS-2035688.}
}%

\author{
	Guangchun~Ruan,~\IEEEmembership{Member,~IEEE,}
	Zekuan~Yu,
	Shutong~Pu,
	Songtao~Zhou,
	Haiwang~Zhong,~\IEEEmembership{Senior~Member,~IEEE,}
	Le~Xie,~\IEEEmembership{Fellow,~IEEE,}
	Qing~Xia,~\IEEEmembership{Senior~Member,~IEEE,}
	Chongqing~Kang,~\IEEEmembership{Fellow,~IEEE}
}

\maketitle

\begin{abstract}
	Intervention policies against COVID-19 have caused large-scale disruptions globally, and led to a series of pattern changes in the power system operation.
	Analyzing these pandemic-induced patterns is imperative to identify the potential risks and impacts of this extreme event.
	With this purpose, we developed an open-access data hub (COVID-EMDA+), an open-source toolbox (CoVEMDA), and a few evaluation methods to explore what the U.S. power systems are experiencing during COVID-19. These resources could be broadly used for research, public policy, and educational purposes.
	Technically, our data hub harmonizes a variety of raw data such as generation mix, demand profiles, electricity price, weather observations, mobility, confirmed cases and deaths. 
	Typical methods are reformulated and standardized in our toolbox, including baseline estimation, regression analysis, and scientific visualization. Here the fluctuation index and probabilistic baseline are proposed for the first time to consider data fluctuation and estimation uncertainty. 
	Based on these, we conduct three empirical studies on the U.S. power systems, and share new solutions and unexpected findings to address the issues of public concerns.
	This conveys a more complete picture of the pandemic's impacts, and also opens up several attractive topics for future work.
	Python, Matlab source codes, and user manuals are all publicly shared on a Github repository.
\end{abstract} 


\section{Introduction} \label{sec:intro}

\subsection{Background}
\IEEEPARstart{T}{he} COVID-19 pandemic is a once-in-a-century crisis for the globe, causing 288.7 million infections and 5.4 million deaths until the end of 2021~\cite{dong2020interactive}.
Governments worldwide took a wide range of non-pharmaceutical interventions in response to the pandemic~\cite{hale2021global}, and as a result, these restrictions and lockdowns have significantly changed the electricity consumption patterns, and had a domino effect through the entire power systems.
Although the power sector has long prepared against a few predictable threats~\cite{wormuth2020electric}, such kind of large-scale, long-term, and high-intensity interference is still quite special.

A systematic~\cite{jiang2021impacts} and empirical~\cite{ruan2021covanl} perspective is critical to understand the pandemic's impacts on power systems. 
For power system operators, COVID-19 has opened up an opportunity to assess the abnormal operation patterns, and identify the future pathways for sustainable recovery~\cite{kanda2020what}.
Existing works have touched part of these topics, but there are still challenges in resource availability and method effectiveness.

This motivates us to record the potential data resources during COVID-19, and develop a support toolbox with adequate built-in methods and metrics (both existing and newly-developed ones) for different groups of people, such as scholars, policy makers, educators, students, and the general public. 
For illustration, our work is expected to promote (i)~research works on power system operation, resilience, or other cross-domain topics, (ii)~policy impact evaluation and empirical analysis, (iii)~adaptive method designs to relieve the pandemic impacts, (iv)~university courses and open webinars to raise public awareness.

The above idea came true as a joint project initiated in May 2020 by Texas A\&M University and Tsinghua University. We soon received a lot of constructive feedback from the power community and industrial partnership, and until now, our work has successfully supported several research teams, university courses, analytical reports, and open webinars.

\subsection{Progress and Literature Review}
Recent empirical studies worldwide have increased our understanding by the observations in France~\cite{obst2021adaptive}, Italy~\cite{gallo2021effects}, Great Britain~\cite{badesa2021ancillary}, the U.S.~\cite{ruan2020covem}, and Canada~\cite{abu2020analysis}. There are extensive literature evaluating the potential impacts in different energy topics, such as power system operation~\cite{agdas2020impact}, household electricity consumption~\cite{zanocco2021exploring}, gasoline demand~\cite{ou2020machine}, energy security~\cite{graff2020covid}, green recovery~\cite{steffen2020navigating}, and climate change~\cite{gillingham2020short}. Taken together, all these efforts provide consistent evidences that COVID-19 has made a series of impacts on bulk power systems~\cite{navon2021effects} as well as mid- to small-scale systems~\cite{carmon2020readiness}. However, the full picture is still unclear up to now, and further efforts are needed to explore the highly coupled influences~\cite{elsarague2022impacts} and adaptive response actions~\cite{luo2022flexibility} in more detail. 

An obvious bottleneck is that the energy community has made very limited efforts to standardize the pandemic-related data and models. While those findings on a case-by-case basis, e.g., \cite{obst2021adaptive,gallo2021effects,badesa2021ancillary}, are still informative, it remains highly complicated to reproduce a published work, or make meaningful comparisons among different results (even for the same country). Even a few efficient methods or models have not been fully understood and applied.

Further demonstrations are given below:

\subsubsection{Data Issue}
Many data resources are not available for the public, especially the cleaned or fine-tuned data. In this case, the similar but tedious data prepossessing could repeatedly dominate the research time of everyone, or even worse, potential academic communication might be interrupted or blocked.
Reference~\cite{obst2021adaptive} collected the power consumption and meteorological data from the French system operator RTE and Meteo-France, but their cleaned dataset was not shared to the public.
Similarly, reference~\cite{gallo2021effects,badesa2021ancillary,abu2020analysis} did not directly share their datasets either.
Up to now, two of the most popular data sources are the U.S. Energy Information Administration (EIA)~\cite{agdas2020impact}, and the European Network of Transmission System Operators for Electricity (ENTSO-E)~\cite{halbrugge2021how}. But users are still required to get familiar with the complex data category and storage rules, and implement all the data preprocessing steps by themselves.
Additionally, it would be tougher to get access to some other rare data sources, e.g., the Indian electricity market data~\cite{kumar2020experience}, or the Swedish building standards and statistics~\cite{zhang2020preliminary}.

Another finding is that most scholars (including the above) have rarely expanded their data category to consider some cross-domain data which may inspire interdisciplinary studies.

\subsubsection{Model Issue}
Quite different methods, models, and criteria are applied in different publications, but very few of them provides an open-source license. 
Benchmarking is so challenging in this condition that one may take a long time to realize even a basic function. 
This is, of course, not friendly to the public, students, and scholars in other fields.
For example, an ordinary least squares model was used in \cite{zanocco2021exploring} to analyze the online survey data in California. Since a few household characteristics and respondent demographics were mentioned, it would need extra efforts to specify the detailed expressions. 
Then a join-point regression was applied in \cite{delgado2021trend} to assess the electricity load trends in Brazil and its geographic regions, but the discussion about model details was somehow limited.
Reference~\cite{norouzi2021impact} developed three time-series models to determine the impacts on the Spain electricity market, and reference~\cite{werth2021impact} used a five-year moving average method to establish a non-pandemic scenario.
It is a pity that both works~\cite{norouzi2021impact} and \cite{werth2021impact} didn't share the codes for public use.

Another trend is that machine learning approaches become increasingly popular in analyzing the potential impacts on the operation or resilience of power systems~\cite{ruan2021mlopt}. 
Reference~\cite{gulati2021impact} used five classical machine learning approaches for electric load forecast in India.
Reference~\cite{zhan2021random} established a random-forest-bagging and board learning system for estimating the daily confirmed cases.
Learning models of other kinds were also effective, including deep learning models~\cite{jamshidi2020artificial}, capsule networks~\cite{saif2021capscovnet}, and domain adaptation~\cite{xu2021cross}. Although powerful, these models made it tough to reproduce a reported result because of their increasing complexity~\cite{mohammed2020benchmarking}.

There is also an urgent need for methodological developments in different use cases. For example, reliable baselines are critical for many applications, but limited contributions are made in stabling the estimation or considering uncertainty problems~\cite{huang2021prediction}. This might become more challenging when our focus is zooming into the sub-categories, such as buildings~\cite{jogunola2022energy} or electric vehicles~\cite{shahriar2022impacts}, which have more complex operational dynamics. It has always been a difficult task to understand the causality in fluctuating variables such as prices, and our empirical knowledge might be limited. New methods are expected to validate and demonstrate any unconventional factors~\cite{ghazani2022nexus}, and avoid degrading the data granularity in distributional calculations like~\cite{ghiani2020impact}.

\subsubsection{Open Source Efforts}
Open source community has actively involved in combating COVID-19~\cite{frazer2020involvement}. Perhaps the most prominent efforts in tracking the pandemic's impacts and sharing open data are made by Johns Hopkins University~\cite{dong2020interactive} and Oxford University~\cite{hale2021global}. In reference~\cite{dong2020interactive}, an interactive dashboard was developed for all affected countries in real time. And an Oxford COVID-19 government response tracker (OxCGRT) was established in \cite{hale2021global} to assess the policy responses of over 180 countries and subnational jurisdictions.

Other efforts include CovidCounties (a public health data tracker at the level of U.S. counties)~\cite{arneson2020covidcounties}, COVID-ResNet (a radiography scanner)~\cite{farooq2020covid}, and OpenABM (an agent-based model for non-pharmaceutical interventions)~\cite{hinch2021openabm}. All of these works, however, were mainly conducted in the public health field, with special focus on the confirmed cases, deaths, government responses, and so on.

One of the few examples from the energy community is reference~\cite{lopez2020impact}, where the authors have made both their data and codes available on Github. This is a positive step forward, but these resources were only collected for five months without timely updates. Dynamic data aggregation is generally needed, but the resources are preferred to be fully free for use, different from the NRGStream (a charged service) in~\cite{leach2020canadian}.

To the best of our knowledge, we are the unique team that develops and constantly upgrades the open-source resources (both data, methods, and toolbox) to track the pandemic's impacts on power systems. Not to mention that we have extensively collected the cross-domain data for interdisciplinary studies.

\subsection{Contributions and Paper Structure}
This paper has made a special effort to evaluate the COVID-19 impacts on power system operation and resilience. Here, the major contributions of our work are summarized as follows:
\begin{itemize}
\item The proposed data hub and toolbox have unique values for data-driven analysis on power system operation and resilience. A variety of data categories from power sector to public health are collected, dynamically updated, and quality-controlled by a support team. The toolbox has two standalone versions developed in Python and Matlab to support diverse users, such as scholars, policy makers, and educators.

\item Typical methods are collected and fully standardized to be more applicable for a wide range of use cases. The fluctuation index and probabilistic baselines are proposed for the first time to consider data fluctuation and confidence interval in baseline estimation. More detailed dynamics and uncertainty features can be effectively detected.

\item Three empirical studies are conducted on U.S. power systems to share several new perspectives, solutions, and unexpected findings. This reflects the high potential value and broad applicability of the proposed resources.
\end{itemize}

Note that we have established a Github repository~\cite{LINK} to launch our data hub and toolbox online before finalizing this paper. This repository has attracted a special attention from the power community, and supported over 40 research groups or individuals up to now, such as a team from Florida State University and New York University~\cite{ospina2021feasibility}. In addition, it has been successfully applied in two graduate courses at Texas A\&M University and Tsinghua University. Some analytical reports and open webinars have featured and introduced our work as well.

The remainder of this paper is organized as follows: Section~\ref{sec:framework} introduces the overall framework, main features, and several quick start guides. Section~\ref{sec:model} demonstrates the details of data, models, and algorithms, then Section~\ref{sec:implementation} discusses the implementation issues in Python and Matlab. Three empirical studies are conducted in Section~\ref{sec:case}. At last, Section~\ref{sec:conclusion} gives the concluding remarks.

\section{Framework} \label{sec:framework}

\subsection{Overall Workflow}
This paper creates a Github repository~\cite{LINK} that consists of an open-access data hub (COVID-EMDA+) and an open-source toolbox (CoVEMDA). One can access these resources from the directories of ``data\_release/'' and ``toolbox/'' respectively.
Note that COVID-EMDA+ is the abbreviation for ``Coronavirus Disease -- Electricity Market Data Aggregation+'', and CoVEMDA for ``CoronaVirus -- Electricity Market Data Analyzer''.

\begin{figure}[t]
	\centering
	\includegraphics[width=0.44\textwidth]{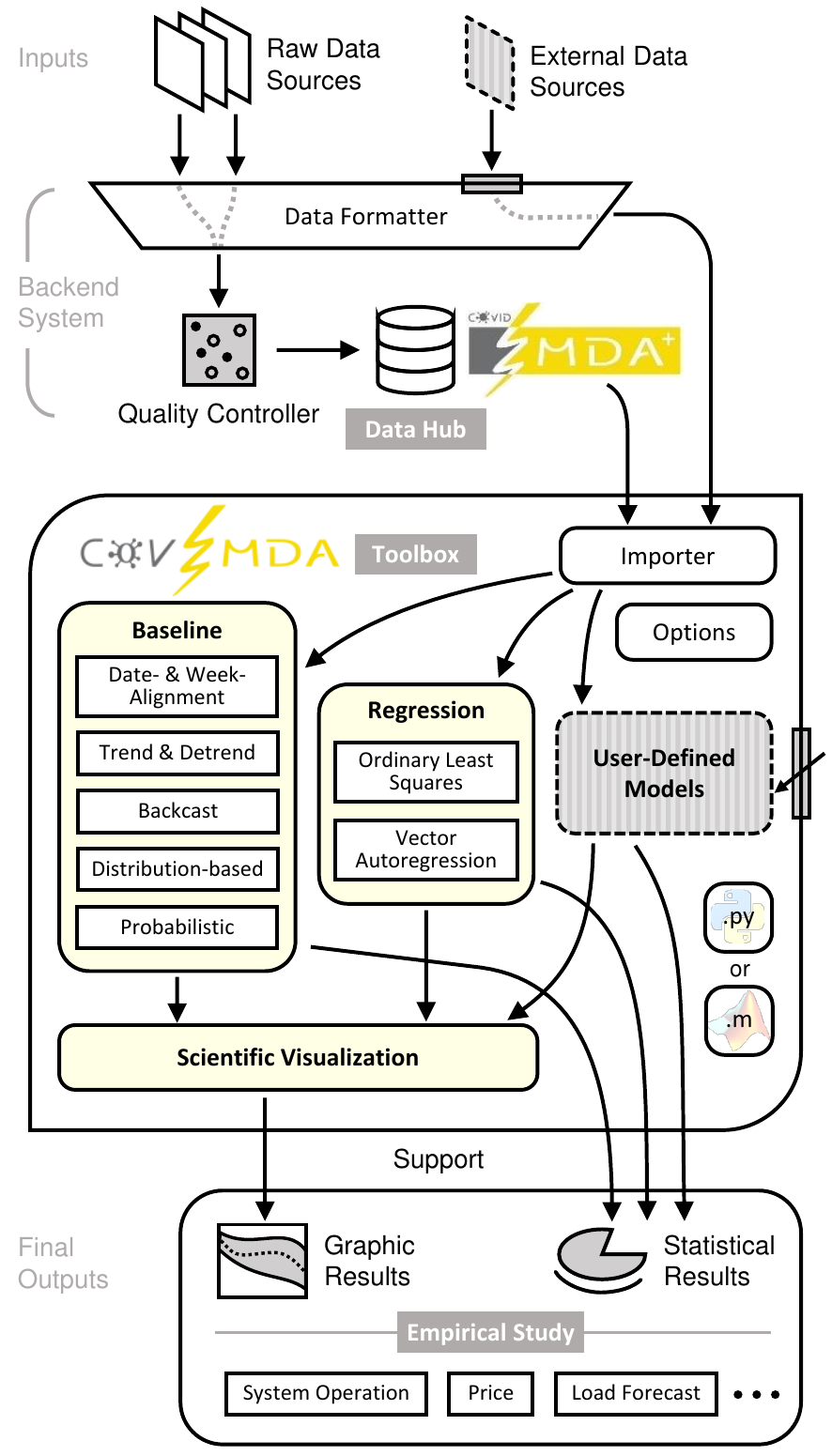}
	\caption{Overall workflow for the proposed data hub and toolbox. All the processing steps from inputs to outputs are shown, and the main functions and extensions are demonstrated as well.}
	\label{fig:framework}
\end{figure}

Fig.~\ref{fig:framework} demonstrates the entire framework and workflow of the proposed data hub and toolbox. 

As shown, the backend system will routinely run the data formatter and quality controller to update the data hub. Outliers and missing data are largely handled with backup data or historical trends, while we also prepare a data quality report to record those highly problematic data.

Fig.~\ref{fig:framework} has listed out three built-in functions in the toolbox: baseline estimation, regression analysis, and scientific visualization. Users are allowed to run this toolbox with Python or Matlab consoles, and generate a variety of graphic and statistical results for further empirical studies. 

In addition, external data and user-defined models are both supported, providing great flexibility for special or advanced extensions. Readers may refer to the online manuscript for further details and quick start examples.

The whole system, including the data hub and toolbox, is maintained by a support team from Texas A\&M University and Tsinghua University. The routine maintenance includes making regular backups, fixing bugs, handling feedback, upgrading online systems, logging, and so on.

\subsection{Main Features}
We summarize the main features of the data hub (COVID-EMDA+) and toolbox (CoVEMDA) as follows:
\begin{itemize}
	\item Data Resources: Broad data categories for the classical as well as (novel) cross-domain analysis on power systems.
	\item Baseline Estimation: Collect five representative methods, two of them are proposed for the first time. Allow rigorous comparisons among different baselines for various power system measurements.
	\item Regression Analysis: Establish two typical and powerful models with built-in statistical tests. Support multiple kinds of model extensions.
	\item Scientific Visualization: Involve tailored designs for a variety of power system applications. 
	\item Empirical Study: Powerful to handle a few technical challenges, report and interpret new empirical findings.
\end{itemize}

\section{Data, Models, and Algorithms} \label{sec:model}

\subsection{Data Sources}
Our data hub collects raw data from multiple sources: 
(i) electricity data from all regional system operators (e.g., CAISO for California, NYISO for New York) along with backup data from EIA and EnergyOnline company, 
(ii) public health data from Johns Hopkins University, 
(iii) meteorological data from Iowa State University, 
(iv) mobile device location data (mobility data) from Safegraph company, and 
(v) satellite image data from NASA (for visualization only). Readers may find all the detailed links for these sources on Github~\cite{LINK}.

Most data records in our data hub could be expressed by $X_{ymdt}$. Here, $X$ is a placeholder for some variable, and the indices collectively specify a time point---year $y$, month $m$, day $d$, and hour $t$. We often use $X_{ymd}$ or $X_{ym}$ to represent different kinds of mean values, for example:
\begin{align}
	\label{eqn:var-sum}
	X_{ym} = \frac{1}{N_{\{d,t\}}} \sum_{\forall d,t} X_{ymdt}
\end{align}
where $X_{ym}$ denotes the mean value of month $m$ in year $y$. It is derived by averaging $X_{ymdt}$ along the axes $d$ and $t$, and $N_{\{d,t\}}$ denotes an auxiliary number.

\subsection{Data Structure and Preprocessing} \label{subsec:data-struct}
One barrier for merging multiple data sources is the inconsistent data formats and structures, which can be further categorized into: (i)~inconsistent data file format and record frequency, (ii)~different definitions and abbreviations, and (iii)~diverse data quality control policies. This motivates us to clean and standardize those messy data by wide data frames (well-known and efficient).

\begin{figure}[t]
	\centering
	\includegraphics[width=0.42\textwidth]{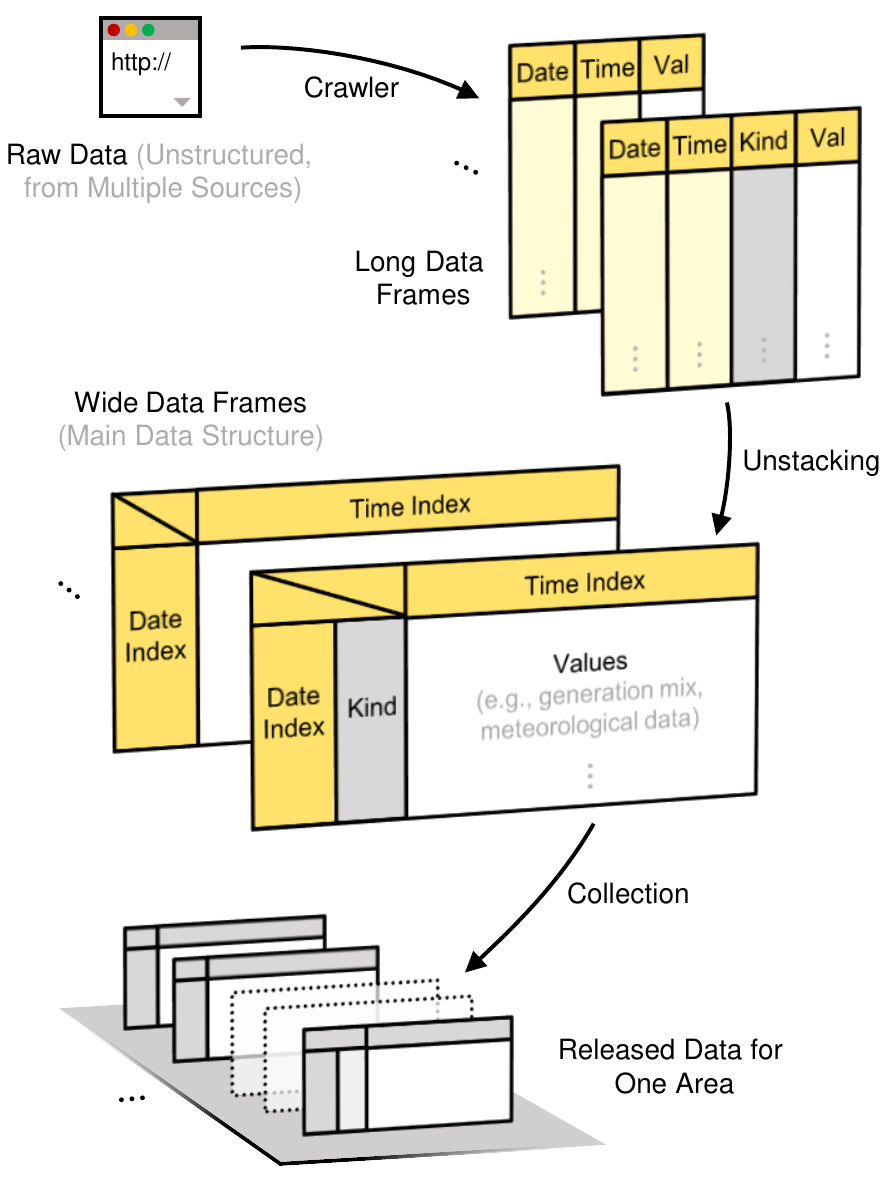}
	\caption{Demonstration of the proposed data structure and preprocessing steps. This procedure is already automated and executed by the backend system.}
	\label{fig:data}
\end{figure}

Fig.~\ref{fig:data} shows the proposed data structure with details. Here, a wide data frame refers to a kind of unstacked table that has more columns than a long frame. This structure enables a more compact way to store data, and both the row-wise and column-wise operations have clear physical meanings. Besides, a variety of basic operational functions (e.g., filtering, resampling, and statistical computing) are available in Python and Matlab to handle such a matrix-like structure.

We store the raw data (e.g., $\{ X_{ymdt} \ \forall y,m,d,t \}$) as a wide data frame by assigning a date index (combining axes $y$, $m$, and $d$) to the rows and an hour index (axis $t$) to the columns.

Fig.~\ref{fig:data} also demonstrates how to finalize the released data after several preprocessing steps. These clean and regularly-updated data can be either downloaded from the Github repository~\cite{LINK} as offline files, or directly retrieved online by using the toolbox functions.

In fact, all the preprocessing steps have been automated by our backend system which consists of a few web crawlers, a set of automation modules, the workflow controller, the quality controller, and a logging module. This backend system is scheduled to run periodically, and for each run, 31 raw data files from 25 sources will be extracted and cleaned to update 73 spreadsheets. Here, outliers and missing data are efficiently detected and handled by analyzing the historical trend or backup data---different rules are specialized for different variables. We further record some problematic data (very rare) in a quality control report for ease of reference.

\subsection{Baseline Estimation} \label{subsec:baseline}
Baselines refer to the reference status for comparison. We focus on estimating a counterfactual baseline that assumes the absence of COVID-19. The difference between a counterfactual outcome and an actual observation will naturally substantiate the pandemic's impacts.

In this aspect, baseline estimation is recognized as the first-and-foremost step for a few impact assessments, and a low-quality baseline may distort our judgment on the influential intensity and duration. 

We next summarize a collection of five built-in methods which are applicable for most applications. The last two methods, proposed for the first time, are effective to consider and inform the uncertainty of estimations.

\subsubsection{Date- and Week-Aligned Estimation}
This method is simple but effective for many use cases when the only or major influential factor is time. The main idea is choosing the proper historical records to be the baselines. 

A date-aligned estimator selects the same date in last year or several years before, shown as:
\begin{align}
	\label{eqn:date-align}
	X_{ymdt} \xrightarrow{\text{ baseline }} X_{y'mdt}
\end{align}
where $y' \le y-1$, and the annotated arrow links an observation (left) with its baseline (right).

A week-aligned estimator selects another historical date which shares the same week-weekday index as the current date. This method is technically formulated as follows:
\begin{align}
	\label{eqn:week-align}
	X_{ymdt} \xrightarrow{\text{ baseline }} X_{y'm'd't}
\end{align}
where the above two dates should satisfy:
\begin{align}
	\label{eqn:week-align-sup}
	f_\text{d2w} (y,m,d) = f_\text{d2w} (y',m',d')
\end{align}

In equation~(\ref{eqn:week-align-sup}), the function $f_\text{d2w} (\cdot)$ calculates the week number and weekday for a given date. For example, $f_\text{d2w} (2020,6,1) = f_\text{d2w} (2019,6,3)$ because both of them are Mondays of the 22nd week.

\subsubsection{Trend and Detrend Estimation}
This method is designed to extract or eliminate the trends' impacts, and thus leads to a better estimation result. Here, the trend can be estimated by either of the following formulas:
\begin{gather}
	T_{ymdt} = f^\text{trend}_w ( X_{ymdt}, \cdots ) \label{eqn:trend-func1} \\
	T_{ymdt} = \hat{f}^\text{trend}_w ( X_{ymdt}, \cdots; \theta^\text{trend} ) \label{eqn:trend-func2}
\end{gather}
where $T_{ymdt}$ is the trend series, $f^\text{trend}_w (\cdot)$ and $\hat{f}^\text{trend}_w (\cdot)$ are two estimation functions, $w$ is a given length of the sliding window, and $\theta^\text{trend}$ denotes the model parameters to be calibrated. 
For illustration, weekly moving average is an instance of (\ref{eqn:trend-func1}), and other advanced models may follow the format of (\ref{eqn:trend-func2}).

A trend and a detrend estimator calculate the baselines differently, shown as follows:
\begin{gather}
	T_{ymdt} \xrightarrow{\text{ baseline }} T_{y'mdt} \label{eqn:trend1} \\
	X_{ymdt} \xrightarrow{\text{ baseline }} T_{ymdt} \label{eqn:trend2}
\end{gather}

The baselines in (\ref{eqn:trend1}) use the trend to remove potential noises, while the baselines in (\ref{eqn:trend2}) detrend the original data to find any additional changes, e.g., extra increments.

\subsubsection{Backcast Estimation}
This method has a complicated expression based on machine learning, so more data and computations are required to calibrate the unknown parameters. This method is originally used to analyze the electricity consumption with great improvement in accuracy. Here, a backcast estimation can be described as follows:
\begin{align}
	\label{eqn:backcast-func}
	B_{ymdt} = \hat{f}^\text{back}_w ( X_{y'mdt}, Y_{y'mdt}, \cdots; \theta^\text{back} )
\end{align}
where $B_{ymdt}$ is the backcast outcome calculated by a machine learning model $\hat{f}^\text{back}_w (\cdot)$, and $\theta^\text{back}$ denotes the corresponding model parameters (often high-dimensional). Here, $X$ and $Y$ represent different input variables.

It is intuitive to extend (\ref{eqn:backcast-func}) to formulate an ensemble backcast model by averaging the outputs of multiple base models (indexed by $i$):
\begin{align}
	\label{eqn:ensemble-backcast-func}
	\hat{f}^\text{back}_w (\cdot) = \frac{1}{N_{\{i\}}} \sum_{\forall i} \hat{f}^\text{back}_{w,i} (\cdot)
\end{align}

Often, a backcast estimation can largely mitigate the adverse impacts of non-pandemic factors to establish a reliable baseline, shown as follows:
\begin{align}
	\label{eqn:backcast}
	X_{ymdt} \xrightarrow{\text{ baseline }} B_{ymdt}
\end{align}

Note that one distinct advantage of this method is the flexibility because there are so many possible options and combinations for the base models.

\subsubsection{Distribution-based Estimation}
This method provides a new distributional perspective to understand the underlying patterns. The key point is monitoring the distributions rather than the raw data when handling some fluctuating variable such as electricity price. A classical metric is given as follows:
\begin{align}
	\label{eqn:distrib-dist}
	S_{ym} = \big\| F_{w,ym} - F_{w,y'm} \big\|
\end{align}
where $S_{ym}$ is a monthly metric of distributional distance. $F_{w,ym}$ and $F_{w,y'm}$ describes the cumulative distribution for month $m$ in year $y$ and $y'$. In both functions, the sliding window $w$ takes the length of one month.

The key disadvantage of $S_{ym}$ is that it significantly reduces the granularity of original time-series data. For instance, using (\ref{eqn:distrib-dist}) will degrade the data granularity from a hourly to a monthly frequency.

To overcome this issue, we develop a novel fluctuation index to capture the distributional features while maintaining the same data granularity. Technically, this index evaluates the possibility that an observation data might be abnormal with the following expression:
\begin{align}
	\label{eqn:index-func}
	I_{ymdt} = f^\text{fluc}_w (X_{ymdt}) = \big| 1 - 2 F_w (X_{ymdt}) \big|
\end{align}
where $I_{ymdt}$ is the proposed fluctuation index, $f^\text{fluc}_w (\cdot)$ is an estimation function that is formulated by the cumulative distribution function  $F_w (\cdot)$. 

\begin{figure}[t]
	\centering
	\includegraphics[width=0.46\textwidth]{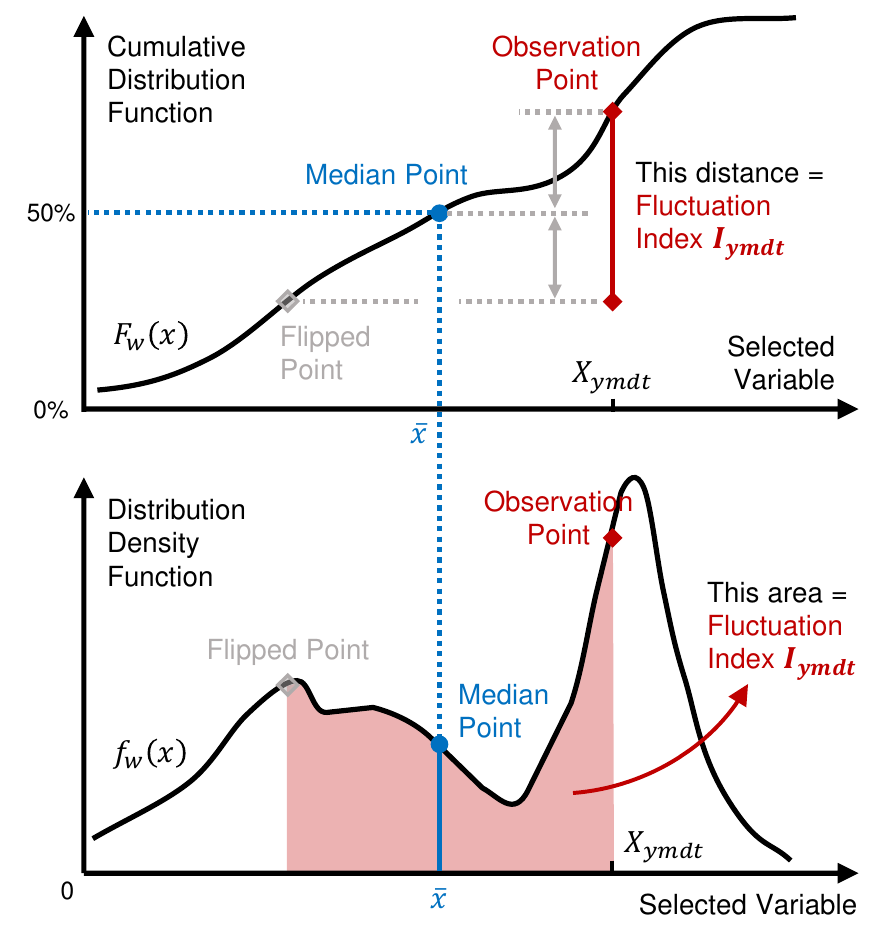}
	\caption{Illustration of the proposed fluctuation index. This index can be physically explained by a highlighted distance in the cumulative distribution curve or a shaded area in the probability density curve.}
	\label{fig:index}
\end{figure}

Fig.~\ref{fig:index} offers a graphic illustration of the fluctuation index from two aspects: the highlighted distance and the shaded area. By definition, $0 \le I_{ymdt} \le 1$, and $I_{ymdt} \ge 0.7$ rarely happens. It is thus possible to evaluate the abnormal dynamics by monitoring this fluctuation index. 

A distribution-based estimator is able to offer a baseline in either of the following way:
\begin{gather}
	I_{ymdt} \xrightarrow{\text{ baseline }} I_{y'mdt} \label{eqn:index1} \\
	I_{ym} \xrightarrow{\text{ baseline }} I_{y'm} \label{eqn:index2}
\end{gather}
where $I_{ym}$ is the monthly average derived similarly as (\ref{eqn:var-sum}), and other possible frequencies are also allowed.

\subsubsection{Probabilistic Baseline Estimation}
This method intends to handle the uncertainty problem in baseline estimation, which is a key knowledge gap that has not been filled by the existing works. In detail, most related works are focused on point estimation, but we extend to consider the baseline uncertainty by developing quantiles in a probabilistic framework, and that also naturally meets the practical needs for constructing confidence intervals.

A probabilistic baseline estimation includes a collection of quantile estimators, each of which should be a parameterized model chosen from the previous subsections, such as:
\begin{align}
	\tilde{T}_{ymdt}(q) = \hat{f}^\text{trend}_w ( X_{ymdt}, \cdots; \theta^\text{trend}_q ) \label{eqn:prob-cal1} \\
	\tilde{B}_{ymdt}(q) = \hat{f}^\text{back}_w ( X_{y'mdt}, \cdots; \theta^\text{back}_q ) \label{eqn:prob-cal2}
\end{align}
where $\tilde{T}_{ymdt}$ and $\tilde{B}_{ymdt}$ are the probabilistic versions of those defined in (\ref{eqn:trend-func2}) and (\ref{eqn:backcast-func}), given $q$ as a quantile value. These models are configured with a pinball loss function to consider the following five indexed quantile levels: $q_1=10\%$, $q_2=25\%$, $q_3=50\%$~(mean value), $q_4=75\%$, and $q_5=90\%$.

The unique feature of this method is that baselines are now estimated by a collection of sub-models, shown as:
\begin{align}
	\label{eqn:prob-baseline}
	X_{ymdt} \xrightarrow{\text{ baseline }} \{\tilde{T}_{ymdt}(q_i), i=1,\cdots,5\} \\
	X_{ymdt} \xrightarrow{\text{ baseline }} \{\tilde{B}_{ymdt}(q_i), i=1,\cdots,5\}
\end{align}

This is more informative than any single deterministic estimator, and one can further establish a 50\% confidential interval by picking out the results of $q_2$ and $q_4$, or a 80\% confidential interval by $q_1$ and $q_5$.

\subsection{Regression Analysis}
Regression is widely used in empirical analysis to explore the potential relationship between different factors. In particular, regression allows us to answer a few questions on correlation or causality during COVID-19. We have collected two popular regression models, along with several useful statistical tests.

\subsubsection{Ordinary Least Squares Regression~(OLS)}
This method offers multiple expressions to check the underlying correlation or causality. Supported formulations include linear expressions as well as a few nonlinear expressions with quadratic, interaction, or logarithms terms. 

An OLS model can be formulated as follows: 
\begin{align}
	\label{eqn:ols}
	Z_{ymdt} = \theta^\text{ols}_1 X_{ymdt} + \theta^\text{ols}_2 Y_{ymdt} + \cdots + \epsilon^\text{ols}_{ymdt}
\end{align}
where $X$, $Y$, $Z$ are placeholders for a group of correlated variables, $\theta^\text{ols}_1$, $\theta^\text{ols}_2$ denote the regression coefficients, and $\epsilon^\text{ols}_{ymdt}$ represents the error term. 

Note that (\ref{eqn:ols}) is highly flexible to represent a series of possible formulations. For example, if all the correlated variables share the same time index, the regression is simply considering the correlation between variables instead of time. Temporal coupling can be considered by adding some historical items. Also, the ellipsis mark in (\ref{eqn:ols}) indicates that other regression terms (linear or nonlinear) are fully allowed.

We calibrate an OLS model by determining a set of regression coefficients to minimize the regression residuals. Here is the related optimization problem:
\begin{align}
	\label{eqn:ols-solve}
	\min \!\! \sum_{\forall y,m,d,t} \!\!\!\!
	\big( Z_{ymdt} - \theta^\text{ols}_1 X_{ymdt} - \theta^\text{ols}_2 Y_{ymdt} - \cdots \big)^2
\end{align}

After calibration, an OLS model can be further validated by running a few statistical tests, including t-test, F-test, and normality-test. R-squared and adjusted R-squared are also informative to evaluate the goodness of fit.

\subsubsection{Vector Autoregression~(VAR)}
This method is specialized to capture the complicated correlation between multiple time-series data. One can extend this method to restricted vector autoregression when some regression coefficients are imposed to be zeros. Both models are powerful and widely adopted in empirical studies. 

A VAR model combines all the variables together and uses the following formula to model the evolution over time:
\begin{align}
	\label{eqn:var}
	X_{ymdt} = \sum_{i=1}^p \theta^\text{var}_i X_{y,m,d,t-i} + \theta^\text{var}_0 + \epsilon^\text{var}_{ymdt}
\end{align}
where $X_{ymdt}$ should be interpreted as one variable or a concatenation of several variables. $p$ is called the order of this VAR model, and the lag terms for the last $p$ periods are considered. Besides, $\theta^\text{var}_0, \cdots, \theta^\text{var}_p$ are regression coefficients, and $\epsilon^\text{var}_{ymdt}$ denotes the error term. 

The establishment of a VAR model can be divided into four steps: pre-estimation preparation, model calibration, model verification, and post-estimation analysis.

First, we need to conduct an Augmented Dickey-Fuller (ADF) test, a cointegration test, and a Granger causality test to analyze the conditions of stationarity, cointegration, and potential causality respecitvely. 

Second, the regression coefficients can be determined by a series of minimization problems, each of which is similar as (\ref{eqn:ols-solve}). For a $p$-order VAR model (\ref{eqn:var}), one should run a total of $p$ optimizations.

Third, another ADF test is used to test if the residual series is stationary, while a Ljung-Box test and a Durbin-Watson test are used to inspect the underlying endogeneity and autocorrelation. A robustness test is then preferred to demonstrate the model performance against coefficient perturbations.

Finally, the calibrated VAR model can provide further insights by running the impulse response analysis and forecast error variance decomposition. 

Note that any other regression models, beyond the above two, can be implemented and used as toolbox extensions.

\subsection{Scientific Visualization}
Scientific visualization is one of the most intuitive way to exhibit empirical findings, but the methods turn out to be highly diverse in different applications. We thus specialize the methods for several classical use cases.

A line chart and a scatter chart are useful to show a series of changing data, such as the raw data $X_{ymdt}$, any aggregated data like $\sum_t X_{ymdt}$, and any filtered data like $X_{ymdt} (m \le 6)$. When the x-axis represents dates, our toolbox further supports highlighting the dates of big events during COVID-19.

A stacked bar chart is able to make comparisons between different categories. Visually, different bars (representing those categories) are stacked end-to-end and assigned different colors for distinction. Assume the raw data $X_{ymdt}$ can be divided into several sub-categories $X_{ymdt}^k \ \forall k$, then the corresponding proportion for $X_{ymdt}^k$ is calculated as:
\begin{align}
	\label{eqn:stack-chart}
	X_{ymdt}^{k\%} = \frac{X_{ymdt}^k}{X_{ymdt}} \times 100\%
\end{align}

A histogram describes the distribution or frequency features of a group of fluctuating data. This is helpful to handle a large amount of observations and detect any possible outliers. In particular, our toolbox supports visualizing the cumulative distribution function and the probability density. 

A box plot is designed to graphically display groups of data through their quantiles. It can effectively handle a data matrix by calculating the quantiles for each column and visualizing these quantiles with color bands. The following five quantiles are of interest:
\begin{align}
	\label{eqn:boxplot}
	Q_i = F^{-1} (q_i), \quad i=1,\cdots,5
\end{align}
where $F (\cdot)$ denotes the cumulative distribution function for one column. $q_1$--$q_5$ have been defined before, just after (\ref{eqn:prob-cal2}).

A heat map is often used to show how a series of data are clustered or varying over space. In such a figure, it is thus easy to understand the spatial correlations, and discover any typical pattern visually as well. Our toolbox implements the two dimensional heat map, and each element is measured by the Pearson correlation coefficient.

It should be noted that a few more visualization schemes can be taken into consideration as toolbox extensions.

\section{Implementation} \label{sec:implementation}

\subsection{Architecture Design} \label{subsec:archit}
We get started by a discussion about the high-level architecture to implement the models and algorithms in Section~\ref{sec:model}. Both standalone versions of our toolbox, developed by Python or Matlab, follow this architecture.

\subsubsection{Folder Structure}
\begin{figure}[t]
	\centering
	\includegraphics[width=0.48\textwidth]{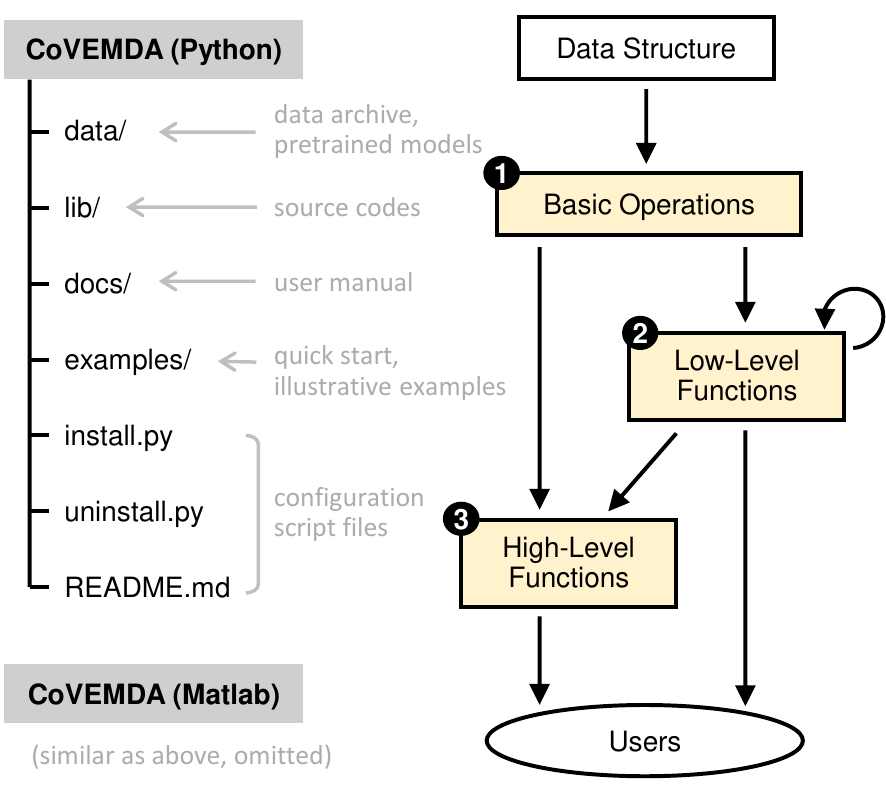}
	\caption{Folder structure (left) and three-level programming architecture (right) for the proposed toolbox.}
	\label{fig:struct}
\end{figure}

Fig.~\ref{fig:struct} shows a concise folder structure in the left part. Note that all the archived data and pre-trained model are prepared in the ``data/'' folder, and the source codes can be found in the ``lib/'' folder. Beginners may get started by reading the user manual or quick start examples. 

\subsubsection{Programming Structure} 
Fig.~\ref{fig:struct} also illustrates a programming structure (on the right) that classifies the entire function family into three levels: basic operations, low-level functions, and high-level functions. It breaks down large tasks (user-oriented) into small activities (data-oriented), and helps clarify the calling relationships and dependencies between different functions.

\subsection{Python Implementation} \label{subsec:python-imp}

\subsubsection{Data Structure}
The toolbox establishes a new DataFormatter class to realize the wide data frame structure mentioned in Subsection~\ref{subsec:data-struct}. This class wraps the popular DataFrame class from Pandas package, and extends the built-in function family with a lot of specialized functions.

\begin{figure}[t]
	\centering
	\includegraphics[width=0.48\textwidth]{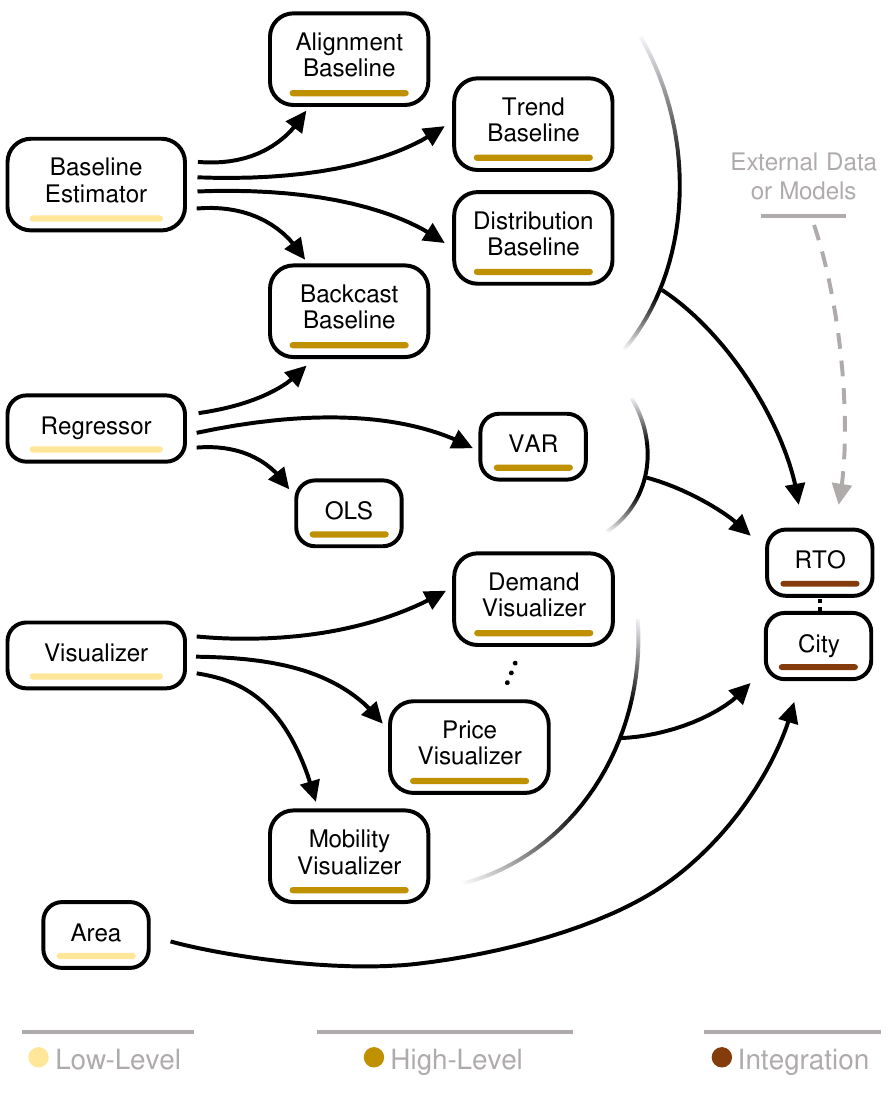}
	\caption{Class inheritance map of the proposed toolbox (Python version). Different classes are designed with functions of different levels. External data and models are also supported for further extensions.}
	\label{fig:class}
\end{figure}

\subsubsection{Object-Oriented Design}
Fig.~\ref{fig:class} elaborates how to organize the major classes and their inheritance relationship to realize the proposed methods. There are four base classes---a baseline estimator class, a regressor class, a visualizer calss, and an area class---they mainly build up the fundamental properties and some key components. A few high-level classes are then established to specify the algorithm details, and finally integrated into the RTO and City class for ease of use.

As for extensions, users are allowed to develop their own class based on the predefined classes. External data sources, special parsers, and user-defined functions could be included in this new class to support further development.

We follow the folder structure in Subsection~\ref{subsec:archit} to organize the Python script files (.py files). Relevant classes and functions are collected in the same file for readability.

\subsection{Matlab Implementation} \label{subsec:matlab-imp}

\subsubsection{Data Structure}
The toolbox constructs a new data structure based on the built-in table array in Matlab. The built-in functions are largely extended to realize all basic operations as planned.

\subsubsection{Functional Design}
Functions are carefully assigned to different abstraction levels (Fig.~\ref{fig:struct}), and most functions share the same or similar names as those in the Python version.

Using the folder structure in Subsection~\ref{subsec:archit}, the Matlab script files (.m files) are collected in three different folders according to the function level.

\section{Empirical Studies} \label{sec:case}

\subsection{Issues of Public Concerns}
The COVID-19 pandemic is exacerbating uncertainty and causing a series of unexpected outcomes in power systems. Among all, we typically pick three issues of public concern:
\begin{itemize}
	\item How much and how long has COVID-19 influenced the operation of U.S. power systems?
	\item How were the electricity prices influenced by COVID-19 and the gas price collapse in 2020?
	\item How to adapt the load forecast models to capture the lock-down patterns during COVID-19?
\end{itemize}

Answers to the above issues will deepen our understanding of the operational risks induced by extreme events of such kind. To fill the knowledge gap, this section conduct empirical studies to share and demonstrate several new findings.

\subsection{Issue~1: Pandemic Impact on Steady State of Power Systems}
This issue covers a broad topics but we focus on three specific aspects that have been less discussed.

\subsubsection{Peak Demand Changes among Different Regions}
For a given region, the reduction of peak demand is assessed for each day and averaged for the whole month:
\begin{align}
	\label{eqn:demand-drop}
	\alpha_{ym} = \frac{1}{N_{\{d\}}} \sum_{\forall d} \left( \frac{B_{ymd} - D_{ymd}^\text{peak}}{B_{ymd}} \right) \times 100\%
\end{align}
where the peak demand $D_{ymd}^\text{peak}$ has a backcast baseline $B_{ymd}$, which can be derived by running the pre-trained backcast model (deep learning model) in the toolbox.

Table~\ref{tab:demand-drop} collects the estimation results for seven U.S. marketplaces. MISO (Midcontinent area) and NYISO (New York) are the top two markets that have experienced more than 10\% drop in both April and May. According to the average reduction rates, the situations in June were largely alleviated for all seven regions, but NYISO appeared to recover much more slowly.

\begin{figure}[t]
	\centering
	\includegraphics[width=0.48\textwidth]{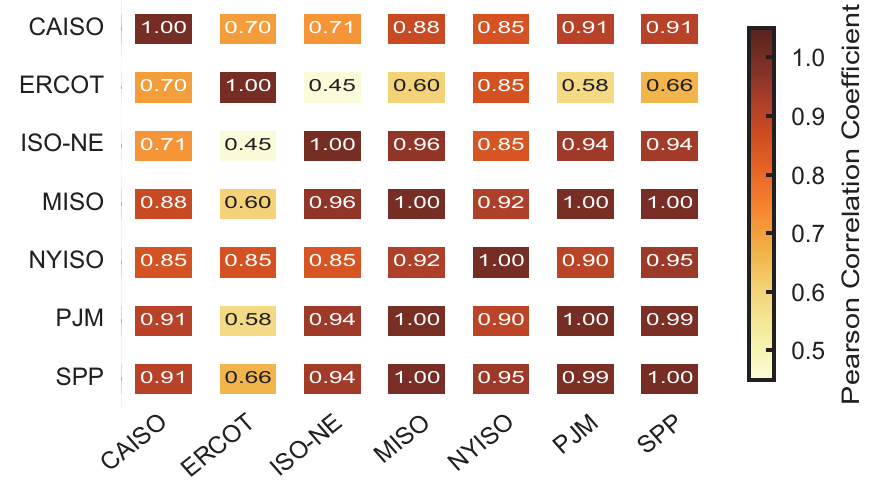}
	\caption{Heatmap of Pearson correlation coefficients between the demand features in different regions. High similarity is expressed by large coefficients and dark colors.}
	\label{fig:heatmap}
\end{figure}

Fig.~\ref{fig:heatmap} then shows the spatial correlation of peak demand changes between different regions. Observing the lightest color in the figure, it is clear to find that CAISO (California) and ERCOT (Texas) exhibits a different pattern, and ERCOT might be the most special one. 

We also calculate the probabilistic baselines in Table~\ref{tab:demand-drop-prob} to understand the situation in ERCOT more clearly. One may check whether the quantiles pass through zero point for probabilistic validation. Here, we are 50\% sure that the peak demand drops from April to June, but it is unclear for March. Larger uncertainty is found in May, 4.71\% within the 25--75\% interval and 8.24\% within the 10--90\% interval, which reflects a underlying pattern change in this month. 

As peak demand is strongly related to the resource adequacy, validating the change and locating the pattern transition could be informative to get the system operator well prepared.

\begin{table}[t] 
	\caption{Reduction Rates of Peak Demand in Different Regions}
	\label{tab:demand-drop} 
	\setlength\tabcolsep{12.3pt}  
	\begin{threeparttable} 
		\begin{tabular}{ccccc}
			\toprule
			Region  & March & April & May   & June \\
			\midrule
			CAISO   & 2.90\%  & \phantom{0}9.28\%  & \phantom{0}6.23\%  & 3.56\% \\
			ERCOT   & --0.85\%\phantom{+} & \phantom{0}3.36\%  & \phantom{0}2.52\%  & 2.70\% \\
			ISO-NE  & 3.14\%  & \phantom{0}6.76\%  & \phantom{0}9.07\%  & 2.33\% \\
			MISO    & 2.57\%  & \textbf{10.23\%} & \textbf{10.70\%} & 2.49\% \\
			NYISO   & \textbf{4.38\%}  & 10.21\% & 10.46\% & \textbf{7.06\%} \\
			PJM     & 1.71\%  & \phantom{0}9.52\%  & \phantom{0}9.08\%  & 1.14\% \\
			SPP     & 0.91\%  & \phantom{0}7.16\%  & \phantom{0}7.08\%  & 1.43\% \\
			\midrule
			Average & 2.11\%  & \phantom{0}8.07\%  & \phantom{0}7.88\%  & 2.96\% \\
			\bottomrule
		\end{tabular}
		\begin{tablenotes}
			\item Note: the largest changes among all the regions are highlighted.
		\end{tablenotes}
	\end{threeparttable}	 
\end{table}

\begin{table}[t] 
	\caption{Probabilistic Reduction Rate of Peak Demand in ERCOT}
	\label{tab:demand-drop-prob} 
	\setlength\tabcolsep{12.6pt}  
	\begin{tabular}{ccccc}
		\toprule
		Quantile & March   & April  & May     & June    \\
		\midrule
		10\%     & --4.01\% & 0.03\% & --1.44\% & --0.17\% \\
		25\%     & --2.53\% & 1.83\% & \phantom{+}0.20\%  & \phantom{+}1.37\%  \\
		75\%   & \phantom{+}2.80\%  & 6.72\% & \phantom{+}4.91\%  & \phantom{+}4.82\%  \\
		90\%   & \phantom{+}4.70\%  & 8.80\% & \phantom{+}6.80\%  & \phantom{+}5.42\%  \\
		\bottomrule
	\end{tabular}	 
\end{table}

\subsubsection{Price Distribution Shift in Chicago}
We then apply the fluctuation index to evaluate the price distributions in Chicago. 

Results show that the monthly index values in 2020 are 0.80~(March), 0.85~(April), 0.62~(May), 0.63~(June); and values in 2019 are 0.46~(March), 0.45~(April), 0.35~(May), 0.52~(June). The largest difference between 2019 and 2020 lies in April when the index value nearly doubles in 2020. Furthermore, the average index value has grown 65.70\%, from 0.44 in 2019 to 0.72 in 2020. These results give reliable evidence that Chicago has experienced a period of severe price changes during COVID-19.

We pay further attention to the weekly statistics for more detailed information of the changing dynamics.  Table~\ref{tab:weekly-price} detects the extreme prices every week that are beyond $2-\sigma$ interval (with the fluctuating indexes beyond 0.9544) and calculates the total number of these prices. Note that the week alignment rule is applied here.

As shown in Table~\ref{tab:weekly-price}, 2020 has witnessed more extreme prices in March and April, and the situation largely relieves after mid-May, but soon rebounds in June. Average number of the extreme price occurrence indicates that it is more than five times more likely to experience an abnormal price (often unexpected low price) in 2020 than 2019.

System operators and power plant managers should take care of these price collapses which may affect the profitability of gas-fired power plants.

\begin{table}[t] 
	\caption{Number of Extreme Price Occurrence in Chicago}
	\label{tab:weekly-price} 
	\setlength\tabcolsep{14pt}  
	\begin{tabular}{ccccc}
		\toprule
		Month & Week & 2019 & 2020 & Increment \\
		\midrule
		March & W1   & 23   & \textbf{38}   & 15        \\
		March & W2   & \phantom{0}1    & \textbf{35}   & 34        \\
		March & W3   & \phantom{0}7    & \textbf{24}   & 17        \\
		March & W4   & \phantom{0}3    & \textbf{62}   & 59        \\
		\midrule
		April & W1   & \phantom{0}2    & \textbf{59}   & 57        \\
		April & W2   & \phantom{0}0    & \textbf{61}   & 61        \\
		April & W3   & \phantom{0}0    & \textbf{39}   & 39        \\
		April & W4   & \phantom{0}4    & \textbf{40}   & 36        \\
		\midrule
		May   & W1   & \phantom{0}0    & \textbf{53}   & 53        \\
		May   & W2   & \phantom{0}0    & \textbf{\phantom{0}9}    & \phantom{0}9         \\
		May   & W3   & \phantom{0}0    & \phantom{0}0    & \phantom{0}0         \\
		May   & W4   & \phantom{0}0    & \phantom{0}0    & \phantom{0}0         \\
		May   & W5   & \textbf{\phantom{0}3}    & \phantom{0}0    & --3        \\
		\midrule
		June  & W1   & 21   & \textbf{39}   & 18        \\
		June  & W2   & 29   & \textbf{56}   & 27        \\
		June  & W3   & 11   & \textbf{44}   & 33        \\
		June  & W4   & \phantom{0}9    & \textbf{31}   & 22		   \\
		\midrule
		Average & & 6.65 & \textbf{34.71} & 28.06 \\
		\bottomrule
	\end{tabular}
	\begin{tablenotes}
		\item Note: Increment value is calculated as the difference between 2020 and 2019. We highlight larger numbers by comparing the numbers in 2019 and 2020.
	\end{tablenotes}
\end{table}

\subsubsection{Duck Curves and Renewable Energy Share in California}
A duck curve, also known as the residual demand, is derived by calculating the difference between electricity consumption and the solar generation:
\begin{align}
	\label{eqn:duck-curve}
	R_{ymdt} = D_{ymdt} - G_{ymdt}^\text{solar}
\end{align}

\begin{figure}[t]
	\centering
	\includegraphics[width=0.45\textwidth]{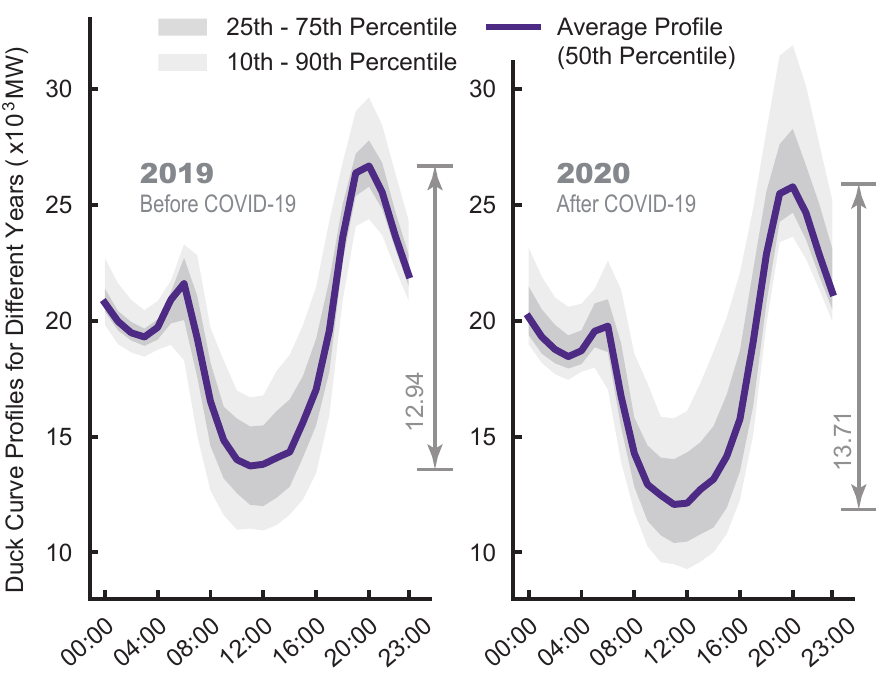}
	\caption{California's duck curves in 2019 and 2020. The ramping is highlighted and labeled with specific numbers.}
	\label{fig:duck-curve}
\end{figure}

Fig.~\ref{fig:duck-curve} compares the duck curves in California. Comparing with 2019, the average duck curve in 2020 has a higher ramping requirement of 761.90 MW, and a larger fluctuation range of 3923.24 MW. As shown, the increased peak-valley difference or peak-valley ratio will call for more flexible resources for power and energy balancing, and system operators should pay close attention to this new situation.

The share of renewable energy is calculated as follows:
\begin{align}
	\label{eqn:renewable-share}
	\beta_{ym} = G_{ym}^\text{hydro\%} + G_{ym}^\text{solar\%} + G_{ym}^\text{wind\%}
\end{align}

We typically consider the monthly proportions in California, and apply an ARIMA model for trend estimation. This model is configured by grid-searching the best hyper-parameters, and the best configuration turns out to be ARIMA(2,0,1). Results show that the observed share of renewable energy during March--June is 34.88\% on average, while the ARIMA model estimates a slightly larger baseline of 34.90\%. This tiny difference, much less than the demand drop, is clearly against the statement that renewables might enjoy extra benefits during COVID-19 because of their low marginal costs. A possible explanation for this finding is the conservative dispatch strategies that take the system safety into consideration.

\subsection{Issue~2: Factor Analysis on Electricity Price Changes}
There is an open debate on the underlying drivers of the low electricity prices during COVID-19, because the pandemic shares a time overlap with the gas price collapse in 2020. Taking Boston as an example, the Pearson correlation coefficient between electricity price and gas price is 0.213, and the coefficient between electricity price and confirmed case number is --0.187, both statistically significant. Spurious correlation may happen in this case, and we have to conduct rigorous regression analysis to figure out the potential causality.

The first step is selecting proper variables and data for the electricity prices and pandemic situations. We calculate the logit value of the fluctuation index, denoted by $LoI_{ymd}$, to describe the abnormality of electricity price observations:
\begin{align}
	LoI_{ymd} = ln \left( \frac{I_{ymd}}{1 - I_{ymd}} \right)
\end{align}
where $I_{ymd}$ is defined similarly as (\ref{eqn:index-func}). Note that $LoI_{ymd}$ has no lower and upper bounds (important for a unbiased OLS regression), and a smaller $LoI_{ymd}$ means more normal. In practice, we regard $0.75 \le LoI_{ymd} \le 3$ as unusual, and $LoI_{ymd} > 3$ as highly unusual.

We also need to construct a gas price variable $\lambda_{ymd}^\text{gas}$ by importing and organizing the data from an external source.
As for the pandemic modeling, we come up with two ways: one is the logarithm of daily confirmed cases $C_{ymd}$, and the other is a binary dummy variable $\delta_{ymd}$ that indicates the absence ($\delta_{ymd}=0$) or presence ($\delta_{ymd}=1$) of the pandemic.

With the above variables, two OLS regression models are designed as follows:
\begin{gather}
	LoI_{ymd} = ( \theta_1 \delta_{ymd} + \theta_2 ) \lambda_{ymd}^\text{gas} + ( \theta_3 \delta_{ymd} + \theta_4 )
	\label{eqn:price-reg1} \\
	LoI_{ymd} = \theta_5 \lambda_{ymd}^\text{gas} + \theta_6 C_{ymd} + \theta_7 \lambda_{ymd}^\text{gas} C_{ymd} + \theta_8
	\label{eqn:price-reg2}
\end{gather}

The basic idea for (\ref{eqn:price-reg1}) and (\ref{eqn:price-reg2}) is controlling the effects of gas prices when assessing the pandemic's impacts. We are also curious about the interaction between these two factors.

Table~\ref{tab:price-reg} illustrates the results of model calibration and statistical tests. We highlight four coefficients that are statistically significant: $\theta_1$, $\theta_3$, $\theta_6$, and $\theta_7$.

\begin{table}[t] 
	\caption{Regression Results of Equation~(\ref{eqn:price-reg1}) and (\ref{eqn:price-reg2})}
	\label{tab:price-reg} 
	\setlength\tabcolsep{12.5pt}  
	\begin{threeparttable} 
		\begin{tabular}{ccccc}
			\toprule
			Parameter & Coeff & Std  & t-Test   & p-Value   \\
			\midrule
			\pmb{$\theta_1$} & \textbf{--2.8715} & \textbf{1.083} & \textbf{--2.652} & \textbf{0.009} \\
			$\theta_2$ & \phantom{+}0.8714  & 0.845 & \phantom{+}1.031 & 0.304 \\
			\pmb{$\theta_3$} & \textbf{\phantom{-}5.4063}  & \textbf{2.016} & \textbf{\phantom{+}2.681} & \textbf{0.008} \\
			$\theta_4$ & --0.6941 & 1.624 & --0.428 & 0.669 \\
			\midrule
			$\theta_5$ & \phantom{+}4.4138 & 2.562 & \phantom{+}1.723 & 0.087 \\
			\pmb{$\theta_6$} & \textbf{\phantom{-}2.8960} & \textbf{0.996} & \textbf{\phantom{+}2.907} & \textbf{0.004} \\
			\pmb{$\theta_7$} & \textbf{--1.4503} & \textbf{0.560} & \textbf{--2.591} & \textbf{0.011} \\
			$\theta_8$ & --8.0344 & 4.548 & --1.766 & 0.079 \\
			\bottomrule
		\end{tabular}
		\begin{tablenotes}
			\item Note: ``Coeff'' is the coefficient value, ``Std'' is the standard deviation. The top part shows the results for (\ref{eqn:price-reg1}), and the bottom part for (\ref{eqn:price-reg2}). In addition, we highlight the rows when the corresponding coefficients are statistically significant.
		\end{tablenotes}
	\end{threeparttable}	 
\end{table}

Here, the pandemic's impact is validated to exist according to a strong statistical evidence that $\theta_1$ and $\theta_3$ in (\ref{eqn:price-reg1}) are nonzero---it turns out to be true that $LoI_{ymd}$ is dependent on $\delta_{ymd}$.

In fact, the influence of COVID-19 is critical because $\theta_3 > 1$ and $\theta_6 > 1$, meaning that the abnormality of prices is really sensitive to the pandemic-related variables.

Another finding is that there may exist an offset relationship between the impacts of COVID-19 and gas prices. One supporting evidence is the negative sign of $\theta_1$. This is further validated by (\ref{eqn:price-reg2}) with a negative $\theta_7$. While the impacts of both factors are synergistic rather than additive (because $\theta_7 \neq 0$), it is at least statically clear that COVID-19 have truly caused more abnormal electricity prices (because $\theta_6 > 0$). 

Note that the above findings still hold true when we run further robust tests by variable substitution and data resampling. Policy makers may need to take effective financial actions to help those power companies with revenue loss during this special time.

\subsection{Issue~3: Adaptive Load Forecast Using Mobility Data}
One severe outcome of COVID-19 is the rapid drop of electricity consumption. Even worse, most load forecast models may fail to capture this sudden break caused by the lockdown policy. This calls for an improved forecasting strategy that could quickly adapt to the new situation and make more accurate predictions.  
We will next show that using mobility data to enhance the load forecast models might be an effective solution. 

This case considers the day-ahead hourly load prediction tasks. Three popular models are tested here: neural network (NN), random forest (RF), and support vector machine (SVM). The inputs for these models include calendar variables, meteorological variables, and the previous load. We also grid-search the hyper-parameters for each kind of model carefully. 

In most cases, the above models cannot capture the novel load pattern during COVID-19, so we improve these models in the following two ways: one is fine-tuning the model with new observations, the other is using mobility data to enhance the results. 

Technically, the latter idea can be described as follows:
\begin{align}
	\label{eqn:forecast-mob}
	\hat{D}_{ymdt} = 
	& \hat{f}^\text{pred} ( D_{y,m,d-1,t}, \cdots ; \theta^\text{pred} ) \notag \\
	& + \Delta \hat{f}^\text{enh} ( M_{y,m,d-1,t} ; \theta^\text{enh} )
\end{align}
where the improved result $\hat{D}_{ymdt}$ has an enhanced item $\Delta \hat{f}^\text{enh} (\cdot)$ that takes the previous mobility data as its input. $\hat{f}^\text{pred} (\cdot)$ is exactly the same as the original model, but we avoid listing all inputs here by an ellipsis mark. 

For simplification, we only consider a linear regression formula for $\hat{f}^\text{enh} (\cdot)$, and we calibrate its parameter $\theta^\text{enh}$ by the residual error series during COVID-19 (very few are needed).

We pay attention to the forecast task in New York City on March 21, two weeks after the state-of-emergence order on March 7, 2020. The main focus is on the prediction performance (measured by mean average percentage errors, or MAPE) of different models in the remaining days before mid-2020. We also validate the performance gaps between the normal period (January 1--March 21) and the lockdown period (March 21--June 30).

Table~\ref{tab:load-forecast} gives the comparison results for different models. They are five kinds in total: the original models, the models updated by new observations (denoted by *--Updated), and the adaptive models enhanced by price/confirmed-case/mobility data (denoted by *--Price/*--Case/*--Mobility). Here the models that use price or confirmed case data share a similar formulation as (\ref{eqn:forecast-mob}).

\begin{table}[t] 
	\caption{Mean Average Percentage Errors of Different Load Forecast Models}
	\label{tab:load-forecast} 
	\begin{threeparttable} 
		\begin{tabular}{lccc}
			\toprule
			Models        & Normal Period & Lockdown Period & Improvement \\
			\midrule
			NN            & 3.10\%        & \phantom{0}8.63\% & \phantom{+}0.00\% \\
			NN--Updated   & ---           & \phantom{0}8.73\% & --0.10\% \\
			NN--Price     & ---           & \phantom{0}6.22\% & \phantom{+}2.41\% \\
			NN--Case      & ---           & \phantom{0}7.34\% & \phantom{+}1.29\% \\
			\textbf{NN--Mobility}  & ---           & \textbf{\phantom{0}5.25\%} & \textbf{\phantom{+}3.38\%} \\
			\midrule
			RF            & 2.84\%        & \phantom{0}8.20\% & \phantom{+}0.00\% \\
			RF--Updated   & ---           & \phantom{0}8.20\% & \phantom{+}0.00\% \\
			RF--Price     & ---           & \phantom{0}6.58\% & \phantom{+}1.62\% \\
			RF--Case      & ---           & \phantom{0}8.93\% & --0.73\% \\
			\textbf{RF--Mobility}  & ---           & \textbf{\phantom{0}6.15\%} & \textbf{\phantom{+}2.05\%} \\
			\midrule
			SVM           & 4.54\%        & 10.79\% & \phantom{+}0.00\% \\
			SVM--Updated  & ---           & 10.69\% & \phantom{+}0.10\% \\
			SVM--Price     & ---           & \phantom{0}7.27\% & \phantom{+}3.52\% \\
			SVM--Case      & ---           & \phantom{0}6.88\% & \phantom{+}3.91\% \\
			\textbf{SVM--Mobility} & ---           & \textbf{\phantom{0}6.84\%} & \textbf{\phantom{+}3.95\%} \\
			\bottomrule
		\end{tabular}
		\begin{tablenotes}
			\item Note: ``NN'' is neural network, ``RF'' is random forest, ``SVM'' is support vector machine. For each part, take the original NN/RF/SVM model as a baseline to calculate the improvement. We also highlight the best estimators of each part.
		\end{tablenotes}
	\end{threeparttable}	 
\end{table} 

It may not be surprising that the performance gaps between the normal and lockdown periods are exceeding 5\%, and some errors are almost tripled. Also, there is nearly no difference when fine-tuning these models with new observations, e.g., RF and its updated model RF-Updated has the same error estimation of 8.20\%.

The major message from Table~\ref{tab:load-forecast} is that using mobility data might improve the forecast performance with an accuracy increase of nearly 25--40\% or 2--4 percentage points. This result can be further improved when obtaining more abnormal observations or increasing the enhancement model size.

Another interesting finding is the best improvements of mobility data integration. In all tests, the mobility-enhanced models work better than those enhanced by price or confirmed case data. Take neural networks as an example, mobility enhancement reduces the error level by 3.38\%, which is 0.93\% and 2.09\% higher than the other two. It can be concluded that mobility is a good and stable indicator for tracking the change of electricity consumption, better than the price or confirmed case data.

Note that the above solution could be useful for system operators to effectively improve their load forecasting models. Our idea, mainly derived from a cross-domain data perspective, is compatible with any model-side improvements to enjoy extra performance benefits.

\section{Concluding Remark} \label{sec:conclusion}
Evaluating the COVID-19 impacts on real-world power systems is critical to understand the potential risks as well as the abnormal operation patterns. But up to now, there is still a lack of reliable and ready-to-use data, methods, and toolboxes for empirical studies.

This paper overcomes the above difficulty by developing an open-access data hub, an open-source toolbox, and a full collection of methods for users with diverse backgrounds, such as scholars, policy makers, and educators. The toolbox is implemented in Python and Matlab with three key functions: baseline estimation, regression analysis, and scientific visualization. The fluctuation index and probabilistic baseline are highlighted because they are general and powerful to provide reliable estimations. We also conduct a few empirical studies with practical evidences to demonstrate new findings and methodologies on three issues of public concerns. 

Typical uses case of the proposed data and toolbox include but not limited to: researches on power system operation and resilience, cross-domain study in energy-related topics, causality analysis on pandemic-induced risks, method designs for adaptive forecasting, policy impact evaluation, and relevant university courses or webinars.

However, our methodologies and toolbox do not involve many specialized methods for spatial correlation analysis, because geological information often lacks documentations and appears in messy formats. Instead, we mainly focus on the temporal relationship, which is often informative enough for pandemic-related problems.

Since the world is confronting a uncertain future induced by COVID-19, this paper may hopefully advance our understanding of the ongoing situations, and guide the preparations through this difficult time.

\bibliographystyle{ieeetr}
\bibliography{refs}

\end{document}